\documentclass{article}
\usepackage{spconf,amsmath,graphicx, amssymb, mathrsfs, subfig, multirow}

\usepackage{enumitem}
\setlist{nosep, leftmargin=14pt}
\usepackage{adjustbox}
\usepackage{ragged2e}
\usepackage{graphicx}
\usepackage{url}
\usepackage{mwe} 
\usepackage{threeparttable} 
\usepackage[normalem]{ulem}
\useunder{\uline}{\ul}{}
\title{
Denoising Diffusion Medical Models
}
\name{
\parbox{\linewidth}{\centering
Pham Ngoc Huy$^{\dagger}$, 
and Tran Minh Quan$^{\dagger\ddagger}$ \\
Email : \{hpham, qtran\}@talosix.com 
}%
}%

\address{
\parbox{\linewidth}{\centering
$^{\dagger}$Talosix, Ho Chi Minh City, Vietnam \qquad
$^{\ddagger}$VinUniversity, Ha Noi, Vietnam 
}
}

\begin{document}
%
\maketitle
\begin{abstract}
In this study, we introduce a generative model that can synthesize a large number of radiographical image/label pairs, and thus is asymptotically favorable to downstream activities such as segmentation in bio-medical image analysis.
Denoising Diffusion Medical Model (DDMM), the proposed technique, can create realistic X-ray images and associated segmentations on a small number of annotated datasets as well as other massive unlabeled datasets with no supervision.
Radiograph/segmentation pairs are generated jointly by the DDMM sampling process in probabilistic mode. 
As a result, a vanilla UNet that uses this data augmentation for segmentation task outperforms other similarly data-centric approaches. 
\end{abstract}
\begin{keywords}
Image Synthesis, Generative Models, Denoising Diffusion, NeRP, ChestXR
\end{keywords}
\section{Introduction}
\label{sec:intro}
X-Ray (XR) images play a critical role in medical diagnosis, especially in the early stage of detecting many diseases~\cite{pmid31636130, pmid28842721}. 
Deep learning applications for XR image processing have evolved in the era of artificial intelligence and achieved some remarkable feats, such as automated segmentation, lesion categorization, quantification, diagnosis and treatment.~\cite{MCBEE20181472}. 
%
To gather such high-quality labels in the biomedical area, however, calls for a significant amount of annotated data, which can be expensive and difficult to obtain.
When training a deep learning model on limited data, a variety of data augmentation approaches are used to avoid overfitting~\cite{perez2017effectiveness, Shorten2019}. 
%
For instance, simple image alterations like morphological geometry adjustments can successfully regularize the training process and aid deep learning techniques in yielding meaningful results.
Recently, diffusion-based models emerged as a promising deep learning method for image generation and have demonstrated their advantages in terms of image quality and training stability.

In this paper, we propose a multi-branch model based on the Denoising Diffusion Probabilistic Model (DDPM)\cite{ho2020denoising} and its improved method, the Denoising Diffusion Implicit Model (DDIM)\cite{song2020denoising}, to synthesize more image/label tuples for use in training models for other biomedical imaging tasks such as segmentation.
The generated samples not only improve diversity and photo-realism in several common metrics, but also generalize dataset distributions with minimal supervisions.
%
In addition, the pairs of image/label sampled by DDMM also improve the results of downstream tasks compared to other state-of-the-art generative models.

Our contributions for the proposed method, namely Denoising Diffusion Medical Models (DDMM), are two-fold: 
First, \textbf{(1)} the DDMM model is built from one or more branches of DDPM (radiographs and segmentation branches) that share the same noise scheduler and latent code, which enforce semantic consistency.
Second, \textbf{(2)} each branch in DDMM can use other unlabeled large-scale datasets to increase diversity and generalization.
These settings make DDMM useful for 
\textbf{(a)} generalizing the dataset distribution by simultaneously generating high-quality XR-like images and their annotations, and 
\textbf{(b)} leveraging the synthesized data to improve other image analysis tasks (such as segmentation) or broaden out-domain image-to-image translation. 

\section{Related Work}
\label{sec:related}
\begin{figure*}[h]
    \vspace{-5mm}
    \centering
    \includegraphics[width=\textwidth]{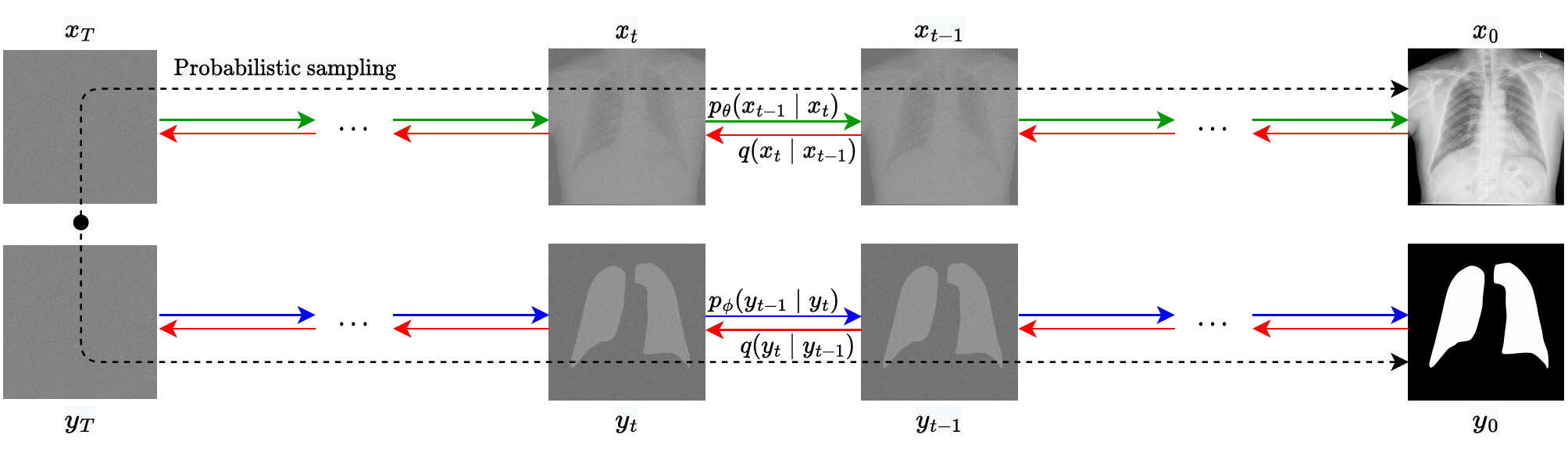}
    \caption{The upper branch is a DDPM model that attempts to denoise the random Gaussian noisy input and produce the XR-like image, while the lower branch tries to generate the corresponding segmentation. 
    To ensure semantical consistency, both use the same initialized noises and noise scheduler.
    }
    \label{fig:model_arch}
\end{figure*}

\subsection{Generative models for XR-like image generation}
XR-like image generation methods were developed using multiple approaches, both physics-based and otherwise. 
The physics-based models, such as XRaySyn~\cite{peng2021xraysyn} or DeepDRR~\cite{unberath2018deepdrr}, produce high-quality images. 
However, their training procedures require knowledge from other modalities, and they can not generate the corresponding segmentation.
In terms of non-physics-based methods, the GAN-based models showed improvements in image synthesis~\cite{cao2020auto}, data augmentation~\cite{shin2018medical}, and style augmentation~\cite{wagner2021structure}. 
However, the GAN approaches often experience unstable training and mode collapse, particularly when generating images from random noise.

Recently, diffusion-based solutions (DDPM~\cite{ho2020denoising}, DDIM~\cite{song2020denoising}, etc.) have emerged as a new method for image synthesis by gradually denoising a random noisy image in large timesteps, during which a temporal encoding is attached to guide a reconstruction UNet~\cite{ronneberger2015u} model. 
These aforementioned problems and innovations motivated us to develop a novel approach of generating paired XR-like images from random noises without facilitating other modality knowledge.

\subsection{Denoising diffusion probabilistic model}

With an input data sample $x_0 \sim q(x_0)$, the forward process $q$ adds the Gaussian noise with variance $\beta_t \in (0, 1)$ at each time-step to the given input $x_{t-1}$ and produces $T$ latents $x_{t}$ where the subscription $t$ ranges from $1$ to $T$:
\begin{align}
    q(x_1,...,x_T | x_0) &:= \prod_{t=1}^T q(x_t | x_{t-1}) \\
    q(x_t | x_{t-1}) &:= \mathcal{N} \left (x_t; \sqrt{1-\beta_t} x_{t-1}, \beta_t \textbf{I} \right )
\end{align}

Supposing that the time-steps $T$ is long enough, and a good beta scheduler is properly designed, the latent $x_t$ approximates a Gaussian distribution. 
Therefore, if the distribution $q(x_{t-1}|x_t)$ is known, we can sample $x_T \sim N(0, \textbf{I})$ and feed it to the forward process to get $q(x_0)$.
On the other hand, a reverse process can be defined as a routing that gradually removes the noise in the inputs, begins at the point $p(x_T) = \mathcal{N}(x_t, \textbf{0}, \textbf{I})$. 
The join distribution $p_{\theta} (x_{0:T})$ is calculated from the starting point by the following Markov chain:

\begin{align}
    p_{\theta} (x_{0:T})~~~ &:= p(x_T) \prod_{t=1}^T p_{\theta} (x_{t-1} | x_t) \\
    p_{\theta} (x_{t-1} | x_t) &:= \mathcal{N} (x_{t-1}; \mu_{\theta}(x_t, t), \Sigma_{\theta} (x_t, t))
\end{align}

In this case, $p$ could be considered an approximation of $q$ in each time step $t$. 
Therefore, $q$ and $p$ are components of a variational auto-encoder. 
The loss can be defined as the variational lower bound on negative log-likelihood, and is formally rewritten as the sum of loss at each step:
\begin{align}
    L_{\text{vlb}} &:= L_0 + \cdots + L_{t-1} + \cdots + L_T
\end{align}
where
\begin{align}
    L_{0~~~~} &:= -\log p_{\theta} (x_0 | x_1) \\
    L_{t-1} &:= D_{KL} (q(x_{t-1} | x_t, x_0) || p_{\theta} (x_{t-1} | x_t)) \\
    L_{T~~~} &:= D_{KL}(q(x_T|x_0) || p(x_T))
\end{align}


\section{Method}
\label{sec:method}

\subsection{Model overview and Training Procedure}

As shown in Fig.~\ref{fig:model_arch}, our DDMM method consisted of \textbf{two} separate DDPM models that shared the noise latent code. 
The model $p_{\theta}$ takes responsibility for generating the XR images while the model $p_{\phi}$ produces the corresponding segmentation. 
The supervised datasets, which include both XR images and labels, are introduced in each training step to compute the reconstruction loss. 
The $p_{\theta}$ branch can be further trained with other unlabeled datasets at the same stage, producing an unsupervised loss. 
The total loss of the training step is the combination of the supervised loss and unsupervised loss. 
While the former loss component drives the model to generate the corresponding pairs of images and labels since both share the same latent noise, the latter loss component supports in expanding data distribution coverage with different sampling points.


\subsection{Probabilistic Sampling Procedures}
Our multi-branch DDMM method can be extended to multi-modal image-to-image translation with a minimal number of paired samples and extensive unpaired images. 
Equivalently, the Gaussian-noise in the latent space plays as a semantic-sharing code across the multiple domains that it can span. 
Therefore, we can perform the sampling procedures in a probabilistic way: we start at one side
of both radiograph/segmentation branches, initialize and fetch the noise into the processes, then gradually denoise them using $p_\theta$ and $p_\phi$ networks. 
This approach results in both image/label tuples generated simultaneously, with one of them satisfying the XR image distribution, while the other one is tightly coupled with its semantic segmentation. 
%
%
With this approach, we can synthetically generate a massive amount of semantically-consistent pairs of images, which is helpful to finetune other downstream tasks.  


\subsection{Implementation Details}
Our DDMM framework is implemented based on the available open source of DDPM~\cite{ho2020denoising} with the learning rate of Adam optimizer set to $1e{-4}$. 
The models $p_\theta$ and $p_\phi$ share the same cosine-based noise scheduler $\beta_t$ but the supervised training noise is initialized with a different seed than the unsupervised scheme. 
We set the total time-steps $T=100$ for all diffusion branches and trained the DDMM for 100 epochs on a workstation equipped with 64 GB of RAM and an NVIDIA GTX 3090 GPU. 
The training took approximately three days to complete. 
%

\section{Data}
\label{sec:data}


\begin{table}[]
\hspace{-7mm}
\caption{Dataset summary. Those which do not have segmentation are used to train the unsupervised branches.}
\centering
\begin{tabular}{|l|c|c|}
\hline
\textbf{Datasets} & \textbf{Images} & \textbf{Annotation} \\ \hline
\hline
ChinaSet \cite{twoChestXray}  & 566 & Lung \\ \hline
Montgomery \cite{twoChestXray} & 138 & Lung \\ \hline
JSRT \cite{jsrtDataset} & 247 & Lung \\ \hline
VinDr-CXR \cite{chestvindr} & 18,000 & N/A \\ \hline
\end{tabular}
\label{tab:datasetStat}
\end{table}

We demonstrated the experiments on the ChestXR images. 
The number of images available in each sub-dataset is shown in Table~\ref{tab:datasetStat}.
There are 951 accompanying images with their lung region segmentations~\cite{twoChestXray}. 
These labeled pairs are split into the training and test sets at a ratio 80:20.
In the ChestXR experiment, we also leverage 15,000 out of 18,000 images from a large-scale public VinDr-CXR dataset~\cite{chestvindr}, which does not have pixel-level segmentation for the lung regions, to adapt on to our unsupervised training step.

\section{Results}
\label{sec:result}
\begin{figure}[]
    \hspace{-3mm}
    \includegraphics[width=\linewidth]{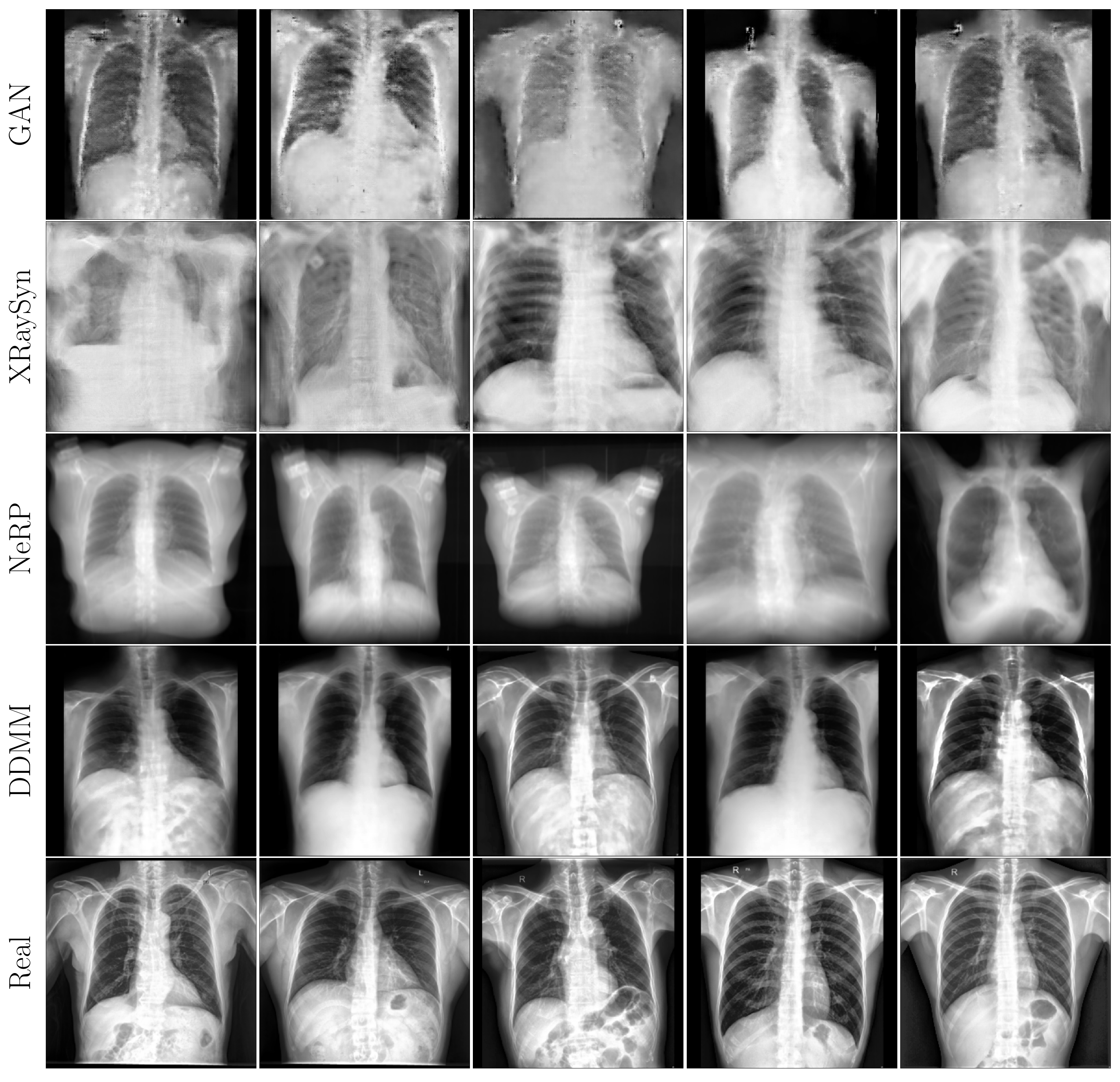}
    \caption{Samples of XR images generated by our method DDMM and others, compared to the real samples in VinDr-CXR test sets (last row).}
    \label{fig:sample_image}
\end{figure}


\begin{table}[]
\hspace{-5mm}
\centering
\caption{Image quality metrics}
\begin{adjustbox}{max width=0.49\textwidth}
\begin{tabular}{|l|c|c|c|c|c|c|}
\hline
\textbf{Method} & \textbf{FID}~$\downarrow$ & \textbf{KID}~$\downarrow$ & \textbf{SSIM}~$\uparrow$ & \textbf{UQI}~$\uparrow$ & \textbf{SCC}~$\uparrow$ \\ \hline
\hline
GAN & 279.869 & 0.3765 & {\ul 0.3724} & 0.0661 & 0.1291 \\ \hline
XRaySyn & 181.390 & 0.2256 & 0.3317 & 0.0318 & 0.0701 \\ \hline
NeRP & 174.294 & {\ul 0.1875} & 0.3268 & 0.0415 & 0.1077 \\ \hline
DDMM (1) & {\ul 155.772} & 0.1913 & 0.3592 & 0.0686 & 0.1325 \\ \hline
DDMM (2) & \textbf{93.998} & \textbf{0.0976} & \textbf{0.4258} & \textbf{0.1012} & \textbf{0.1750} \\ \hline
\end{tabular}
\end{adjustbox}
\captionsetup{labelformat=empty}
\justifying\caption{
\footnotesize
DDMM~(1) is trained with the supervised branch only, while DDMM~(2) is trained with both supervised and unsupervised branches.
}
\label{tab:img_quality}
\end{table}

\subsection{Quality of generated XR images}
Fig.~\ref{fig:sample_image} illustrates the generated samples from DDMM and other methods compared to the actual images. 
The other two approaches, GAN and XRaySyn, produce low-quality images with much distortion and blur. 
The images from NeRP are higher quality, as compared to GAN and XRaySyn, but still miss the bone details in ChestXR. 
These problems are not observed in the DDMM-generated images.
The image quality is further evaluated quantitatively by collecting the numbers of Frechet Inception Distance (FID)~\cite{heusel2017fid}, Kernel Inception Distance (KID)~\cite{binkowski2018kid}, the Structural Similarity Index Measurement (SSIM)~\cite{wang2004ssim}, the Universal Image Quality Index (UQI)~\cite{wang2002universal}, the Spatial Correlation Coefficient (SCC)~\cite{Zhou1998}. 
While the better methods have lower FID and KID scores, the SSIM, UQI, and SCC should be as high as possible.
These metrics are calculated on 1,000 XR images drawn randomly from 10,000 pre-generated images for each method. 
On the ChestXR dataset, the results shown in Table~\ref{tab:img_quality} indicate that DDPM produces the highest quality of synthesized images. 
%
%
In general, without the unsupervised branch, we can not enhance the images. 

\subsection{Image Segmentation}

\begin{table}[h]
    \centering
    \caption{Segmentation results}
    \label{tab:segmentation}
    \begin{tabular}{|c|c|c|c|}
    \hline
    \textbf{Anatomy} & \textbf{Method} & \textbf{Dice Score}~$\uparrow$ & \textbf{Rand Score}~$\uparrow$ \\ \hline
    \hline
    Chest & NeRP & 0.6619 & 0.4057 \\ \hline
    Chest & DDMM & \textbf{0.7649} & \textbf{0.5763} \\ \hline
    \end{tabular}
\end{table}

\begin{figure}[h]
    \centering
    \hspace{-3mm} 
    \includegraphics[width=\linewidth]{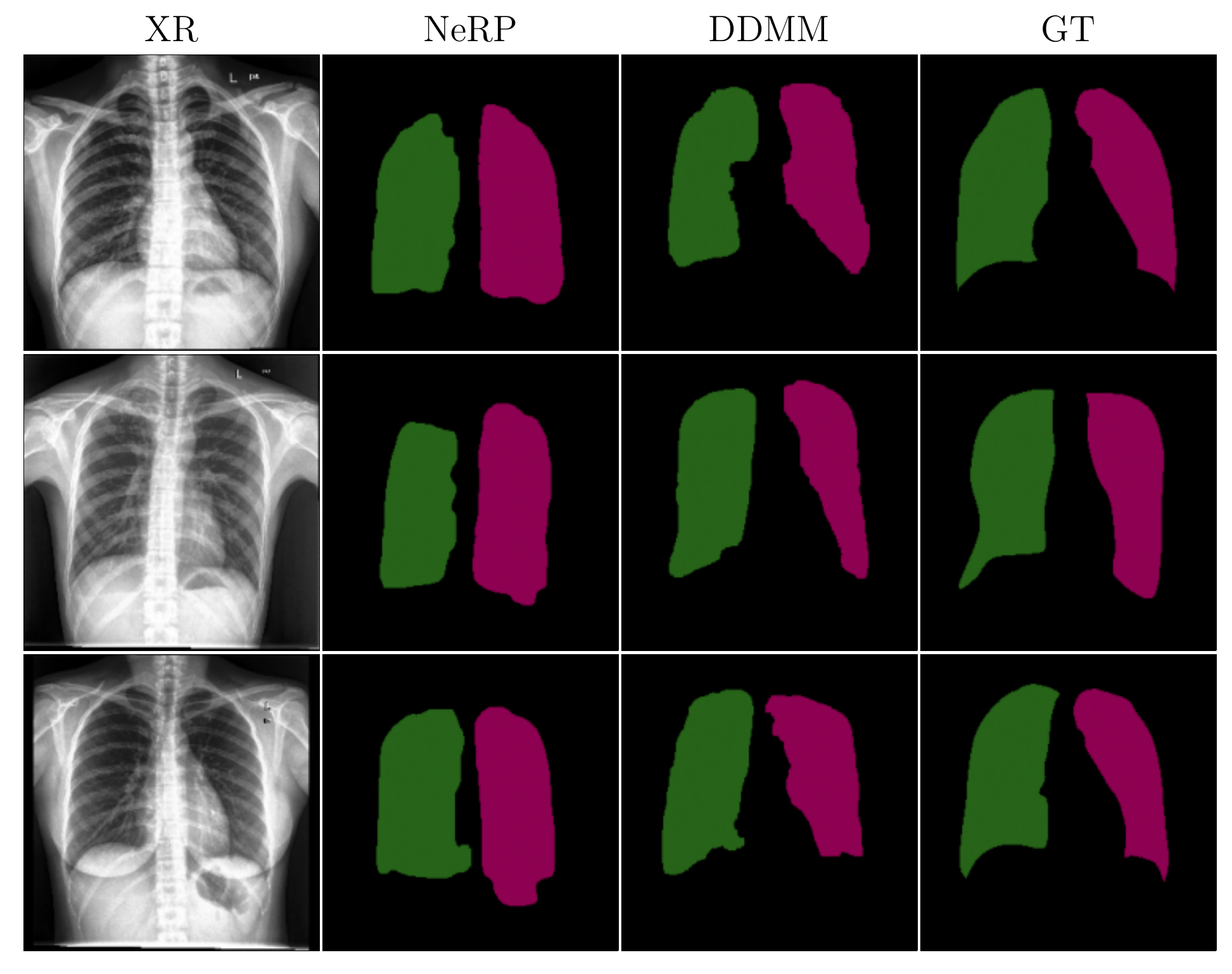}
    \caption{Qualitative evaluations on ChestXR test set of NeRP and DDMM methods against the ground truth.}
    \label{fig:lung_seg}
    \vspace{-3mm} 
\end{figure}

\begin{figure}[h]
    \centering
    \hspace{-3mm} 
    \includegraphics[width=1.0\linewidth]{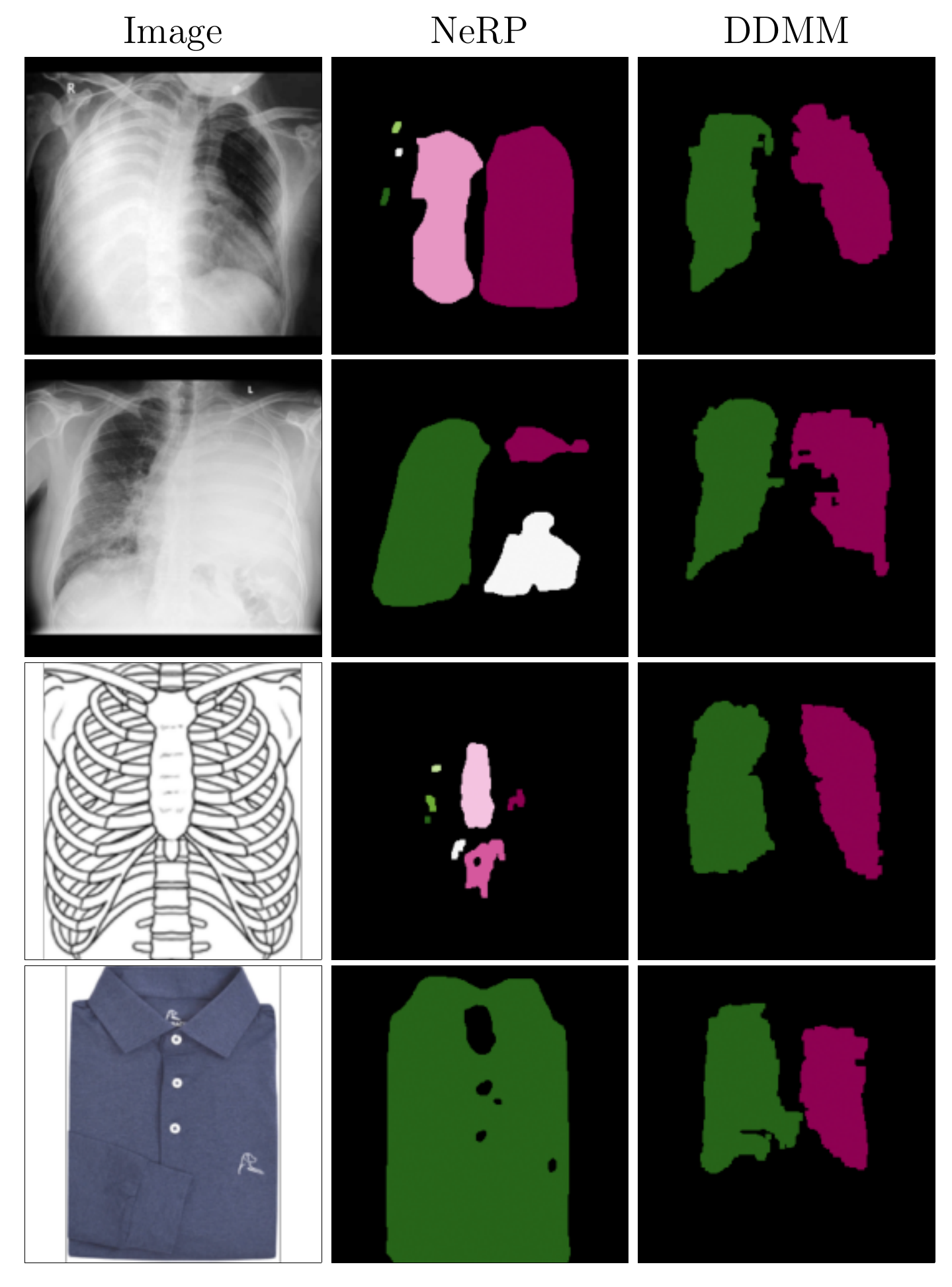}
    \caption{Lung segmentation task on the out-of-domain samples}
    \label{fig:lung_ood}
\end{figure}

\textbf{In-domain segmentation:} 
For each experiment on the above Chest datasets, we pre-generate 10,000 pairs of image/label using NeRP~\cite{nerp2022} and our method DDMM for training vanilla UNets~\cite{ronneberger2015u} without other augmentation techniques. 
We do not reuse the previously split training sets for this downstream task. 
The segmentation performances are assessed directly on the test sets by Dice-score and Rand-score metrics, which measure the semantic-aware and instance-aware significance. 
As can be seen in Fig.~\ref{fig:lung_seg} and Table~\ref{tab:segmentation}, our DDMM method extracts the lung areas qualitatively better than the physics-based method NeRP, and also achieves better segmentation metrics. 

\textbf{Out-of-domain segmentation:}
The segmentation models are tested on the out-of-domain samples. 
The model trained on the DDMM images performed well on the chest XR images even though the input sources are extreme outliers. 
For example, the pleural space with partially visible dark textured lung regions, or even the severe cases of the skeleton and t-shirt images (see Fig.~\ref{fig:lung_ood}), can be inferred more appropriately with our method. 
Interestingly, the model based on DDMM data still detects the lung area on the non-XR images, while the NeRP model cannot produce the proper segmentation. 
These extreme out-of-domain samples showcase the remarkable generalization capability.

\section{Conclusion}
\label{sec:conclusion}
We present DDMM, a diffusion-based multi-branch model that can jointly produce realistic XR medical images and their associated segmentation masks. 
The generated images outperform other similar work qualitatively and quantitatively. 
Our method is also beneficial for downstream tasks, such as improving the segmentation results in biomedical image analysis. 
In addition, DDMM is scalable and can be extended to other cross- and intra-modality such as CT, MRI, or multi spatial omics analysis. 
By leveraging the unlabeled datasets, DDMM better generalizes the data distribution and helps to capture more useful information, which in turn supports better diagnosis, treatment and precision medicine.



\section{Compliance with Ethical Standards}
\label{sec:ethics}

All procedures performed in studies involving human participants were in accordance with the ethical standards of the institutional and/or national research committee and with the 1964 Helsinki declaration and its later amendments or comparable ethical standards.
For this type of retrospective study, formal consent is not required. 

\section{Conflicts of Interest}
\label{sec:conflict}
The authors have no conflicts of interest to disclose. 

\section{Acknowledgment}
\label{sec:cknowledgment}
The authors would like to extend their thanks to Truc Thai, Trista Pham, Dennis Chien, Adam Rhine, and Arsenio Valdez (at Talosix~LLC.) for their invaluable support and contribution towards the methodology and discussion of this paper. 
The authors are truly grateful for the support and assistance provided during the course of this project.

%

\section{References}
\bibliographystyle{IEEEbib}
\bibliography{refs}

\begin{thebibliography}{10}

\bibitem{pmid31636130}
S.~H. Bradley et~al.,
\newblock ``{{S}ensitivity of chest {X}-ray for detecting lung cancer in people
  presenting with symptoms: a systematic review},''
\newblock {\em British Journal of General Practice}, vol. 69, no. 689, pp.
  e827--e835, Dec 2019.

\bibitem{pmid28842721}
F.~Malgo et~al.,
\newblock ``{{V}alue and potential limitations of vertebral fracture assessment
  ({V}{F}{A}) compared to conventional spine radiography: experience from a
  fracture liaison service ({F}{L}{S}) and a meta-analysis},''
\newblock {\em Osteoporosis international}, vol. 28, no. 10, pp. 2955--2965, 10
  2017.

\bibitem{MCBEE20181472}
Morgan~P. McBee, Omer~A. Awan, Andrew~T. Colucci, Comeron~W. Ghobadi, Nadja
  Kadom, Akash~P. Kansagra, Srini Tridandapani, and William~F. Auffermann,
\newblock ``Deep learning in radiology,''
\newblock {\em Academic Radiology}, vol. 25, no. 11, pp. 1472--1480, 2018.

\bibitem{perez2017effectiveness}
Luis Perez and Jason Wang,
\newblock ``The effectiveness of data augmentation in image classification
  using deep learning,''
\newblock {\em arXiv preprint arXiv:1712.04621}, 2017.

\bibitem{Shorten2019}
Connor Shorten and Taghi~M. Khoshgoftaar,
\newblock ``A survey on image data augmentation for deep learning,''
\newblock {\em Journal of Big Data}, vol. 6, no. 1, July 2019.

\bibitem{ho2020denoising}
Jonathan Ho, Ajay Jain, and Pieter Abbeel,
\newblock ``Denoising diffusion probabilistic models,''
\newblock in {\em NeurIPS}, 2020, vol.~33, pp. 6840--6851.

\bibitem{song2020denoising}
Jiaming Song, Chenlin Meng, and Stefano Ermon,
\newblock ``Denoising diffusion implicit models,''
\newblock in {\em ICLR}, 2021.

\bibitem{peng2021xraysyn}
Cheng Peng, Haofu Liao, Gina Wong, Jiebo Luo, S~Kevin Zhou, and Rama Chellappa,
\newblock ``{XraySyn: Realistic View Synthesis From a Single Radiograph Through
  CT Priors},''
\newblock in {\em AAAI}, 2021, vol.~35, pp. 436--444.

\bibitem{unberath2018deepdrr}
Mathias Unberath, Jan-Nico Zaech, Sing~Chun Lee, Bastian Bier, Javad Fotouhi,
  Mehran Armand, and Nassir Navab,
\newblock ``{DeepDRR--a catalyst for machine learning in fluoroscopy-guided
  procedures},''
\newblock in {\em MICCAI}, 2018, pp. 98--106.

\bibitem{cao2020auto}
Bing Cao, Han Zhang, Nannan Wang, Xinbo Gao, and Dinggang Shen,
\newblock ``{Auto-GAN: self-supervised collaborative learning for medical image
  synthesis},''
\newblock in {\em MICCAI}, 2020, vol.~34, pp. 10486--10493.

\bibitem{shin2018medical}
Hoo-Chang Shin~et al.,
\newblock ``Medical image synthesis for data augmentation and anonymization
  using generative adversarial networks,''
\newblock in {\em IWSSMI}, 2018, pp. 1--11.

\bibitem{wagner2021structure}
Sophia~J. Wagner, N.~Khalili, R.~Sharma, M.~Boxberg, C.~Marr, Walter~de Back,
  and T.~Peng,
\newblock ``Structure-preserving multi-domain stain color augmentation using
  style-transfer with disentangled representations,''
\newblock in {\em MICCAI}, 2021, pp. 257--266.

\bibitem{ronneberger2015u}
Olaf Ronneberger, Philipp Fischer, and Thomas Brox,
\newblock ``U-net: Convolutional networks for biomedical image segmentation,''
\newblock in {\em MICCAI}, 2015, pp. 234--241.

\bibitem{twoChestXray}
S.~Jaeger, S.~Candemir, S.~Antani, Y.~X. Wang, P.~X. Lu, and G.~Thoma,
\newblock ``{Two public {CXR} datasets for computer-aided screening of
  pulmonary diseases},''
\newblock {\em Quantitative Imaging in Medicine and Surgery}, vol. 4, no. 6,
  pp. 475--477, Dec 2014.

\bibitem{jsrtDataset}
J.~Shiraishi et~al.,
\newblock ``{Development of a digital image database for chest radiographs with
  and without a lung nodule: receiver operating characteristic analysis of
  radiologists' detection of pulmonary nodules},''
\newblock {\em American Journal of Roentgenology}, vol. 174, no. 1, pp. 71--74,
  Jan 2000.

\bibitem{chestvindr}
Ha~Q. Nguyen, Khanh Lam, Linh~T. Le, Hieu~H. Pham, Dat~Q. Tran, Nguyen, et~al.,
\newblock ``{VinDr-CXR}: An open dataset of chest x-rays with radiologist’s
  annotations,''
\newblock {\em Scientific Data}, vol. 9, no. 1, pp. 1--7, 2022.

\bibitem{heusel2017fid}
Martin Heusel, Hubert Ramsauer, Thomas Unterthiner, Bernhard Nessler, and Sepp
  Hochreiter,
\newblock ``Gans trained by a two time-scale update rule converge to a local
  nash equilibrium,''
\newblock 2017, vol.~30.

\bibitem{binkowski2018kid}
Mikołaj Bińkowski, Dougal~J. Sutherland, Michael Arbel, and Arthur Gretton,
\newblock ``Demystifying {MMD} {GAN}s,''
\newblock in {\em ICLR}, 2018.

\bibitem{wang2004ssim}
Zhou Wang, Alan~C Bovik, Hamid~R Sheikh, and Eero~P Simoncelli,
\newblock ``Image quality assessment: from error visibility to structural
  similarity,''
\newblock {\em IEEE transactions on image processing}, vol. 13, no. 4, pp.
  600--612, 2004.

\bibitem{wang2002universal}
Zhou Wang and Alan~C Bovik,
\newblock ``A universal image quality index,''
\newblock {\em IEEE signal processing letters}, vol. 9, no. 3, pp. 81--84,
  2002.

\bibitem{Zhou1998}
J.~Zhou, D.~L. Civco, and J.~A. Silander,
\newblock ``A wavelet transform method to merge landsat tm and spot
  panchromatic data,''
\newblock {\em International Journal of Remote Sensing}, vol. 19, no. 4, pp.
  743--757, 1998.

\bibitem{nerp2022}
Pham~Ngoc Huy and Tran~Minh Quan,
\newblock ``Neural radiance projection,''
\newblock in {\em IEEE ISBI}, 2022, pp. 1--5.

\end{thebibliography}

\end{document}